\documentclass[bib]{ba}

\usepackage{epsfig}
\usepackage{bayes}
\usepackage{aas_macros}

\begin{document}

\inserttype[ba0001]{article}
\author{Ford, Moorhead \& Veras}{%
  Eric B. Ford\\University of Florida
  \and
  Althea V. Moorhead\\University of Florida
  \and
  Dimitri Veras\\University of Florida \& Institute of Astronomy}
\title[Bayesian Surrogate Model for Time Series Analysis]{A Bayesian Surrogate Model for Rapid Time Series Analysis and Application to Exoplanet Observations}
\maketitle

\begin{abstract}

We present a Bayesian surrogate model for the analysis of periodic or quasi-periodic time series data.  
We describe a computationally efficient implementation that enables Bayesian model comparison.  
We apply this model to simulated and real exoplanet observations. We discuss the results and demonstrate some of the challenges for applying our surrogate model to realistic exoplanet data sets.
In particular, we find that analyses of real world data should pay careful attention to the effects of uneven spacing of observations and the choice of prior for the ``jitter'' parameter.  
\end{abstract}

\section[Motivation]{Motivation \& Overview of Exoplanet Observations}
Since the 1990's, nearly 500 extrasolar planets (or exoplanets) have been discovered around other stars in our galaxy, yet only a few of which have been observed directly.  In all other cases, the planet's presence has been inferred from its influence on the light of its host star.  

\subsection{Doppler Observations}
The most productive such method to date has been observing the Doppler shift of the star light due to the gravitational perturbations of the planet.  For a single planet on a circular orbit, the Doppler signature arises from a sinusoidal variation in the star's velocity along the line of sight.  More generally, a single planet causes a periodic variation in the  stellar velocity  ($v(t)$) that follows the shape predicted by Kepler's laws of planetary motion, which can usually be well-approximated by the first few terms of a Fourier expansion \citep{Konacki99}.  If the amplitude of the fundamental term (with frequency $f=2\pi/P$, where $P$ is the orbital period) in the Fourier expansion of $v(t)$ is $K_0$, then the coefficients of harmonic terms (with frequency $nf$) are of order $e^n K_0$, where $e$ is the orbital eccentricity.  Given the typical signal-to-noise of detections and the typical exoplanet eccentricity of less than 0.3 (and often less than 0.1), the use of as few as two terms in this Fourier expansion is often accurate to within observational uncertainties.  If a star has multiple planets, its Doppler signature can be much more complex.  In some cases, the observed stellar velocities ($v_k$) can be well approximated as the linear superposition of multiple planets on Keplerian orbits.  In other cases, the planet-planet interactions cause effects comparable to the overall amplitude of the signal and thus must be considered.  In any case, the observational signature is quasi-periodic.  A common question is whether a $N_p$ planet model is sufficient to describe the available observations or whether the data demand at least $N_p+1$ planets.  While an exploration of the full physical parameter space would be computationally prohibitive, a lower dimensional surrogate model can be quite useful for analyzing such a system \citep{Veras11}.

\subsection{Transit Timing Variations}
\begin{figure}[ht]
\begin{center}
\epsfig{file=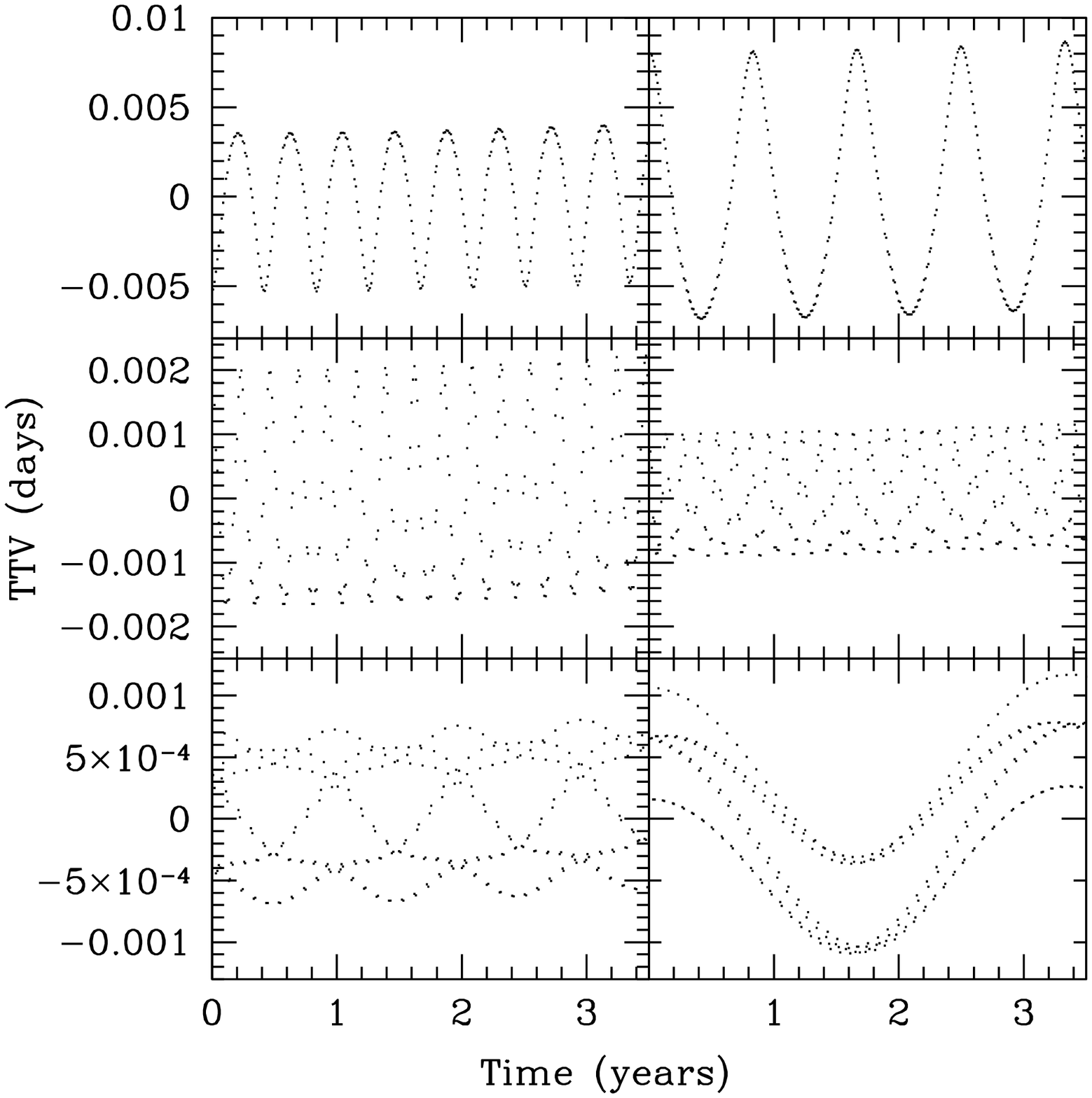,width=4.5in}
\end{center}
\caption{In this figure, we show the transit timing signature for a series of two planet systems, each extending for 3.5 years, the nominal mission lifetime for NASA's {\em Kepler} space observatory.  In each case, transit times are for a 20 Earth-mass planet following an initially circular orbit with a star-planet separation of approximately 0.05AU.  (One AU is the average distance between the Earth and the Sun).  In each case, there is an additional 2 Earth-mass planet (which we assume is not observed to transit) following a slightly eccentric orbit ($e_2=0.1$) and an initial mean star-planet separation of: 0.078AU (upper left), 0.080AU (upper right), 0.080AU (middle left), 0.082AU (middle right), 0.084AU (lower left) or 0.086AU (lower right).  These separations are near the location of the 1:2 mean motion resonance ($\simeq~0.0794$AU).  Note that the vertical axis scale changes from row to row.  Even these small changes in the orbital separation result in qualitative changes in short and long-term structure of the transit timing variations.  Here we show simulated data from full n-body integrations with no data gaps and no measurement uncertainties.  In practice, even {\em Kepler} misses some transits (e.g., due to data downlink with Earth, spacecraft abnormalities) and the transit timing measurements have noise of $\sim~10^{-3}-10^{-2}$ days, depending primarily on the brightness of the target star and the size of the planet.  For some planetary systems \citep[e.g., Kepler-9 b\&c][]{Holman10} or triple star systems \citep{Carter11,Slawson11,Steffen11b}, the amplitude of the transit (eclipse) timing variations is much larger than {\em Kepler's} measurement uncertainties.  For other planetary systems \citep[e.g., Kepler-11 b-f][]{Lissauer11b}, the magnitude of transit timing variations are comparable to the measurement uncertainties.  We expect the Bayesian approach and our surrogate model, to be most useful for such systems, once a sufficient number and timespan of observations have been collected.  For many systems with no detectable transit timing variations, statistical methods such as those described here will play an important role in establishing the significance of non-detection and the implied upper limits for the mass of any perturbing planets \citep[e.g.,][]{Steffen05}.}

\label{fig_ttv_21}
\end{figure}

\begin{figure}[ht]
\begin{center}
\epsfig{file=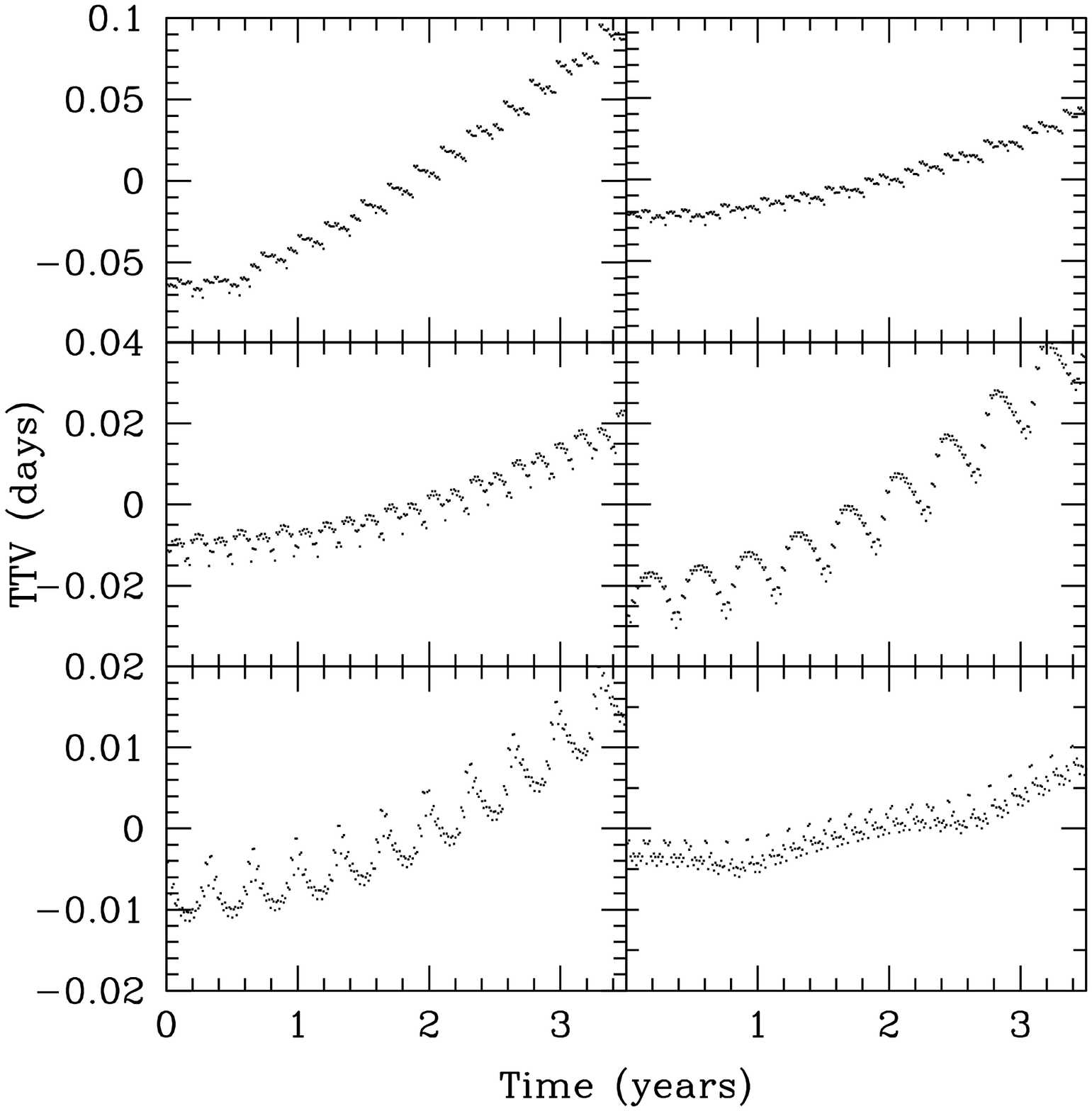,width=5.5in}
\end{center}
\caption{This figure is similar to Figure \ref{fig_ttv_21}, except the the outer planet has been moved closer to the transiting planet.  In each case, there is a 2 Earth-mass planet (which we assume is not observed to transit) following a slightly eccentric orbit ($e_2=0.1$) and an initial mean star-planet separation of: 0.062AU (upper left), 0.063AU (upper right), 0.064AU (middle left), 0.065AU (middle right), 0.066AU (lower left) or 0.067AU (lower right).  These separations are near the location of the 2:3 mean motion resonance ($\simeq~0.0655$AU).  Note that the vertical axis scale changes from row to row.  Again, even these small changes in the orbital separation result in qualitative changes in short and long-term structure of the transit timing variations.}
\label{fig_ttv_32}
\end{figure}

Recently, an extremely promising new method of detecting exoplanets has burst on to the scene.  
As a planet passes in front of its host star (i.e., a transit), the star's brightness appears to decrease.  In an idealized two-body system, the mid-times of the transits ($t_k$) are strictly periodic at the orbital period ($P$), so the $k$th transit occurs at time $t_k = t_0 + k\times~P$.  If there are additional planets, then the times of the transits deviate from a linear ephemeris \citep{Agol05,HolmanMurray05,FordHolman07}.  The perturbations of an additional planet can cause deviations of the transit times that are simply sinusoidal in time, a periodic non-sinusoidal pattern or a very complex quasi-periodic waveform \citep[Figure \ref{fig_ttv_21}, \ref{fig_ttv_32};][]{FordHolman07,Nesvorny08,Nesvorny09,Nesvorny10}.  While it is impractical to explore all the parameters of a full physical model, a lower dimensional surrogate model may be able to help identify regions of parameter space that merit further investigation with a full physical model.

\subsection{Relation to Previous Research}
In this manuscript, we present a new method for analyzing periodic time series data using a computationally efficient Bayesian surrogate model.  
The details of our model are chosen to facilitate the analysis of exoplanet observations.  
We test our model by analyzing Doppler and transit timing data sets.  
Thanks to the computational efficiency of our algorithm, it is possible to apply it to a large library of simulated data sets to understand how the model performs for different types of planetary systems.

Astronomers routinely apply Markov chain Monte Carlo (MCMC) techniques to perform Bayesian parameter estimation when analyzing Doppler observations of an exoplanet host star \citep[e.g.,][]{Ford05,Gregory05}.  For multiple planet systems, MCMC methods are computationally intensive, even when the model evaluation is performed ignoring the gravitational interaction between the planets.  While this is a good approximation for analyzing the Doppler observations of many systems, mutual planetary interactions can be quite significant for planetary systems near a mean motion resonance \citep[e.g., GJ 876;][]{Laughlin05}.  Full n-body integrations to account for these mutual planetary interactions are much more computationally demanding.  One approach is to develop and parallelize computationally efficient methods that allow one to use full n-body integrations \citep[e.g.,][]{JohnsonPayne11}.  
While the GJ 876 data set has many high signal-to-noise observations, most exoplanet host stars have fewer observations and lower signal-to-noise, resulting in weak constraints on many physical model parameters and making it even more difficult to sample these parameter spaces using MCMC.  The use of a lower-dimensional surrogate model has the potential to contribute to the analysis of such systems by identifying the periodicities that are statistically significant without introducing several additional parameters that are poorly constrained.  
Similar methods are routinely applied in a frequentist context to identify planet candidates from Doppler observations \citep{Cumming08}.  
Our Bayesian surrogate model can be thought of as a Bayesian generalization of the Lomb-Scargle periodogram \citep{Cumming04} that has been further generalized to allow for multiple frequencies, perhaps due to the perturbations from additional planets or perhaps due to significant eccentricity \citep{Konacki99}.  Previously, a much more restricted version of the surrogate model ($N_{f,max}=1$, $N_{d,max}=0$, see \S\ref{secModel}) was used to evaluate strategies for scheduling Doppler observations \citep{Ford08}.  The generalization in this manuscript allows for identifying multiple periodicities, as is necessary for application to eccentric and/or multiple planet systems.

We are optimistic that the surrogate model has even more potential for contributing to the analysis of transit timing variation data.  In the transit timing variation method, the entire signal is due to the mutual gravitational perturbations.  Given the highly non-linear nature of the problem (particularly near resonances), a physical model requires performing computationally expensive n-body integrations.  While it might be practical to perform MCMC sampling around one mode of the posterior distribution while using full n-body integrations, it is not practical to perform a brute-force global search of the high-dimensional parameter space while using full n-body integrations \citep{Veras11}.  The evaluation of our surrogate model is orders of magnitude faster than an n-body integration.  Additionally, the surrogate model is linear in most of its model parameters, allowing for efficient identification of the modes and integration over linear parameters, once we condition on the non-linear parameters.  (We perform integration over non-linear parameters via brute force, as described in the supplementary materials.)  The speed of the surrogate model makes it well-suited to exploring a broad range of possible orbital configurations.  Once the surrogate model identifies significant periodicities, n-body integrations can be used to perform a more detailed exploration of the full physical models in regions which could produce the periodicities identified by the surrogate model.  We present results of our Bayesian surrogate model applied to simulated transit timing data and discuss the implications of our results for the prospects of transit timing-based planet searches.

\section{Bayesian Surrogate Model for Analysis of quasi-Periodic Time Series}

\subsection{Data}
First, we describe our general model for analysis of quasi-periodic time series.  In the case of Doppler observations, the independent variable, $x$ would correspond to time ($t$) and $y$ would correspond to the star's velocity ($v(t)$).  In the case of transit timing observations, $x$ would correspond to the transit number and $y$ would correspond to the mid-transit time.  Each observation ($y_k$) is accompanied by an estimate of the measurement uncertainty ($\sigma_k$).  The independent variables (i.e., transit number or time of each observation) are assumed to be known precisely.

\subsection{Model}
\label{secModel}
We explore the use of a surrogate model given by

\begin{equation}
y(x_k; \theta) = \sum_{i=1}^{N_f} \left[ S_i \sin\left(2\pi~f_i x_k\right)+ C_i \cos\left(2\pi~f_i x_k\right) \right] + \sum_{i=0}^{N_d} D_i x_k^i,
\end{equation}
where $x_k$ is the independent variable for the $k$th observation and and $y(x_k;\theta)$ is the predicted value of the observable based on the surrogate model with parameters $\theta$.  
The surrogate model parameters $\theta$ include:  
1) the number of frequencies in the surrogate model ($N_f$), 
2) the frequencies ($f_i$'s), 
3) the amplitudes for those periodicities ($S_i$'s and $C_i$'s),
4) the order of the polynomial terms ($N_d$), and
5) the polynomial coefficients ($D_i$'s).
This model exactly describes time series which are the superposition of polynomial and sinusoidal signals.  The surrogate model can be used for Bayesian model comparison to determine how complex a model (i.e., how many frequencies and/or polynomial terms) is justified for a given data set.

In principle, one could consider alternative basis functions.  We favor the use of sinusoids since they exactly describe the gravitational perturbation of a planet following a circular orbit (and non-interacting with other planets).  Further, a sinusoid often provides a good first approximation to the perturbation of a planet on an eccentric orbit, given typical eccentricities and measurement precision.  
In practice, we will truncate the model to use just a few frequencies and polynomial terms, so as to provide an acceptable model for the observations while facilitating the efficient evaluation of the model.  If observations span a sufficiently long period of time, then there is no need for polynomial terms.  In practice, the orbital period of an outer planet (e.g., Saturn 29.5 years, Neptune 165 years) can be much longer than the time span of observations (typically $\sim$1-10 years).  If the orbital period is much longer than the time span of observations than a simple linear term (constant radial acceleration for Doppler data) is all that can be discerned from the available data.  In the case of a planet on a circular orbit, the ``jerk'' (i.e., derivative of acceleration) becomes significant as the time span of observations approaches a quarter of the orbital period.  For the general case of a planet on an eccentric orbit, the jerk can become evident after a greater or lesser duration, depending upon the orientation of the orbit and phase of the observations.  In either case, we cannot infer the physical parameters with just two derivatives measured, as these are insufficient to characterize an orbit that requres more parameters to fully specify (seven physical parameters in general, and at least three even if we arbitrarily assume a circular, coplanar orbit).  In these cases, using only one or two polynomial terms dramatically increases the computational efficiency, since they eliminate the need to explore a large number of frequencies.  Another advantage of the polynomial basis is that a simple linear term corresponds to the Doppler perturbation from a body sufficiently distant that its orbital motion is insignificant over the time span of observations.  A quadratic term corresponds to physical motion of the perturber over the time span of observations.

In cases where the time span of observations is sufficient to discern a third polynomial term, it becomes more difficult to justify a polynomial model.  On one hand, a cubic term is still much less computationally demanding than considering an additional harmonic term (see \S\ref{secModelEval}).  On the other hand, according to our physical model, all signals are periodic on sufficiently long timescales and three derivatives are sufficient to constrain the range of plausible periods and amplitudes.  While three derivatives are not sufficient to infer the shape of the waveform, in many cases (e.g, a distant planet on nearly circular orbit) a simple sinusoid is a reasonable first approximation.  Therefore, we favor considering a model with an additional harmonic term over including three or more polynomial terms.  

\subsection{Likelihood}
We take the likelihood of each observation to be
\begin{equation}
L_{k}(x_k, y_k, \sigma_k |\theta)=N(y_k-y(x_k;\theta), \sigma_k^2+\sigma_j^2)
\end{equation}
where $N(\mu,\sigma^2)$ is a normal distribution with mean $\mu$ and variance $\sigma^2$.  
Here $\theta = \left\{ N_f, f_1, ... f_{N_f}, S_1, ... S_{N_f}, C_1, ... C_{N_f}, N_d, D_0, ... D_{N_d}, \sigma_j \right\}$.  Note that we have expanded the set of model parameters ($\theta$) to include $\sigma_j$, a ``jitter'' parameter that is related the amount of scatter that is not accounted for by the observational uncertainties.  The origin of the jitter need not be specified.  It could be due to inaccurate estimates of the observational uncertainties or physical effects that are not included in a particular model (e.g., star spots, $p$-modes, additional planets).  

We assume that the observational errors are uncorrelated in time, so the likelihood for a data set of $N_{obs}$ observations is given by
\begin{equation}
L(\theta) = \prod_{k=1}^{N_{obs}}L_{k}(x_k, y_k, \sigma_k | \theta) \, .
\end{equation}

\subsection{Priors}
In principle, one might aim to choose priors that are based on the distribution of masses and orbital periods for exoplanets.  Of course, characterizing those distributions is one of the primary motivations to conduct exoplanet searches.  While exoplanet searches have detected hundreds of planets, most of these have masses comparable to Saturn or greater and orbital periods of a few years or less \citep{Cumming08}.  Inevitably, astronomers pushing the frontiers of knowledge (e.g., searching for less massive planets) will not know the intrinsic distribution of those planets' orbital properties.  Thus, we choose broad priors based on physical intuition and mathematical principles, as outlined in this section.

For the number of frequencies in the surrogate model, we adopt the following prior:
$p(N_f) = (1-2\alpha+\alpha^N_{f,max})/(1-\alpha)$ for $N_f=0$,
$p(N_f) = \alpha^{N_f}$ for $0 < N_f \le~ N_{f,max}$, and
$p(N_f) = 0$ for $N_f > N_{f,max}$,
where $\alpha$ parametrizes our prior belief about the likelihood of multiple frequencies in the signal.
The maximum number of frequencies to be considered ($N_{f,max}\ge~1$) is chosen so as to provide enough complexity to model the data while keeping the model evaluation practical.  
The choice of a geometric probability for an increasing number of frequencies has the advantage that the prior ratio $p(N_f=n+1)/p(N_f=n)$ is independent of $n$ (as long as $n+1\le~N_{f,max}$).  If we associate each frequency with one planet and consider planet detection serially (i.e., first look for evidence of one planet, next look for evidence of another planet), then the minimum Bayes factor to have a significant detection of each successive planet is constant.

The prior for the frequencies themselves are given by
$p(\log f_i) \sim U(\log f_{\min},\log f_{\max})$,
where $U(a,b)$ is a uniform distribution between $a$ and $b$.  For our applications, there are physical limits on the range of viable frequencies.  For example, $f_{\max}$ could be set by the shortest orbital period in which a planet would be able to survive for the age of the star, and $f_{\min}$ could be longest  orbital period in which a planet would be able to remain bound for the age of the star given perturbations from the galactic tidal field and passing stars.  In most cases, the time span of observations ($T_{obs}$) will not be sufficient to distinguish such long-period signals from a low-order polynomial.  Thus, we typically set $f_{\min}\sim~2/T_{obs}$, as the surrogate model is still able to model slow variations and this aids in the rapid evaluation of the surrogate model.  
Our choice of a prior that is flat in the logarithm of the frequency is motivated by the maximum entropy prior for a scale parameter.  We also experimented with a modified Jeffrey's prior.  We found no significant difference in the results, as long as there is a significant detection.  This can be understood simply in terms of the characteristic width of the peaks in the likelihood are much smaller than the domain of the frequencies.  Thus, as long as the prior for frequency has significant support across the entire domain, the choice of the prior for frequency has minimal impact on the shape of the posterior (in locations with significant posterior probability) unless there is not significant evidence for any periodicities.  
While we do not attempt to justify our choice of prior based on the distribution of exoplanet orbital periods, we note that this alternative approach would result in a fairly similar choice of prior, at least for frequencies between $\sim$1/(2 days) and $\sim$1/(2000 days) \citep{Cumming08}.  

Each of the frequencies ($f_i$) has an amplitude $A_i = \sqrt{S_i^2+C_i^2}$ and a phase $\phi_i = \mathrm{atan2}(-S_i,C_i)$.  We choose a uniform prior for each phase, $p(\phi_i) = 1/\left(2\pi\right)$.  Physically, this corresponds to time invariance; i.e., the other planetary system does not know what time we choose to label as $t=0$.  
For $A_i$, the total amplitude at frequency $i$, we adopt a modified Jeffrey's prior, $p(A_i) \sim \left(1+A_i/A_o\right)^{-1}$ for $0\le~A_i\le~A_{\max}$, where $A_{\max}$ is the maximum plausible amplitude.   For application to Doppler observations, $A_{\max}$ could be set by the maximum velocity perturbation by a planet (or binary star).  For application to transit timing variations, $A_{\max}$ could be set by the orbital period.  Note that, in principle, $A_{\max}$ could be a function of $f_i$.  
The other parameter, $A_o$, prevents a divergence of the prior at small amplitudes.  For some applications $A_o$ may be physically motivated.  For our applications, we choose $A_o$ based on the minimum detectable signal based on the available data set, e.g., $A_o\sim~\left<1/\sigma_i^2\right>^{-1/2}$, where $\left<1/\sigma_i^2\right>^{-1/2}$ is effectively the weighted average measurement precision.  
In practice, we find that any sufficiently small choice of $A_o$ and large choice of $A_{\max}$ gives similar results for parameter estimation for a given $N_f$.  The choice of $A_o$, $A_{\max}$ and $\alpha$ do affect the marginalized posterior probability ratio for $N_f$.  

For our typical applications, the total signal amplitude is proportional to the mass of the planet.  
While we do not attempt to justify our choice of prior based on the distribution of exoplanet masses, we note that this alternative approach would result in a fairly similar choice of prior, at least for readily detectable planets (e.g., more massive than Saturn for Doppler surveys).  Present observations are only beginning to provide significant constraints on the distribution of planet masses for low-mass planets at small short orbital periods ($\le$50 days) \citep{Howard10,Borucki11,Youdin11}.  

We implement the above by using a ``two-dimensional modified Jeffrey's prior'' for each pair of amplitudes $S_i$ and $C_i$,
\begin{eqnarray}
\label{EqnSC}
p(S_i,C_i) =1/\left[4\pi \sqrt{S_i^2+C_i^2} \log\left(1+A_{\max}/A_o\right) \left(1+\sqrt{S_i^2+C_i^2}/A_o\right)\right] \hspace{0.5in} && \\ \nonumber
 \mathrm{for} \sqrt{S_i^2+C_i^2}\le~A_{\max} , &&
\end{eqnarray}
and $p(S_i,C_i) = 0$ for $\sqrt{S_i^2+C_i^2}>A_{\max}$.  
Inspection of Eqn.\ \ref{EqnSC} shows that it is essentially a modified Jeffery's prior for the total amplitude $A_i$.  

For the number of polynomial terms in the surrogate model, we adopt a prior similar to that for the number of frequencies:
$p(N_d) = (1-2\beta+\beta^{N_{d,max}-1})/(1-\beta)$ for $N_d=1$,
$p(N_d) = \beta^{N_d-1}$ for $1 < N_d \le~ N_{d,max}$, and
$p(N_d) = 0$ for $N_d > N_{d,max}$, 
where $\beta$ parametrizes our prior belief about the likelihood of higher order polynomial terms being present in the signal.
For our applications, the maximum polynomial order considered ($N_{d,max}\ge~1$) is typically set to 1 and rarely more than 2.  
The choice of a geometric probability for an increasing number of polynomial terms has the advantage that the prior ratio $p(N_d=n+1)/p(N_d=n)$ is independent of $n$ (as long as $n+1\le~N_{d,max}$).  

For both the jitter and the polynomial coefficients, i.e., each of $B \in$ ($D_i$, $\sigma_j$), we adopt a modified Jeffrey's prior,
\begin{equation}
p(B)=1/\left[2 \left|B\right| \log\left(1+B_{\max}/B_o\right) \left(1+\left|B\right|/B_o\right)\right], \qquad \left|B\right|\le~B_{\max},
\end{equation}
where $B_o$ and $B_{\max}$ are analogous to $A_o$ and $A_{\max}$.  We adopt the same values as for $A_o$ and $A_{\max}$.  For some data sets, it may be wise to choose a smaller $B_o$, as the minimum measurable magnitudes for the $D_i$'s depend on the effective measurement precision, the number of observations and the time-span of observations.  In many cases, the amplitude of long-term trends is proportional to the perturbing planet's mass, motivating a prior $p(D_i)$ that is concentrated at small signals and that has a similar shape to the prior for the amplitudes.  We also tried using a uniform prior, $p(D_i) = 1/(2A_{\max})$ for $\left|D_i\right|\le~A_{\max}$.  In practice, the likelihood is very sharply peaked in $D_1$, so results are not sensitive to the choice of its prior.  We have only begun to experiment with the surrogate model for $N_d=2$ (based on discussion at end of \S\ref{secModel}), and caution that further experimentation may be needed for models with $N_d\ge~2$.  

In this manuscript, we present results based on a modified Jeffrey's prior, as the jitter is a scale parameter and we have physical reason to limit $p(\sigma_j)$ at small values of $\sigma_j$, once it becomes small compared to the measurement precision or astrophysical effects that can cause non-gravitational perturbations to the Doppler or transit timing signal (e.g., non-uniform star spots).  However, we found that in some cases our results could be sensitive to the prior for the jitter \citep{Payne11}.  For example, either imposing an upper cutoff on the prior for $\sigma_j$ at $\sim~10$m/s or using a normal distribution for $p(\sigma_j)$  can significantly increase the posterior probability for $N_f=n+1$ relative to that of $N_f=n$ or narrow the range of allowed amplitudes.   Previous studies of the empirical distribution of Doppler jitter are based on relatively small samples sizes \citep{Wright05}, so they have little to say about the tails of the distribution.  Based on our results, we encourage further observations and statistical analyses that could inform the choice of prior for $\sigma_j$.

\subsection{Numerical Evaluation of Model and Posterior PDF}
\label{secModelEval}
The surrogate model was designed to provide a good approximation to Doppler or transit timing observations and is likely to provide a reasonable approximation of many other time series.  A second important feature of the surrogate model is that it permits efficient evaluation.  Several tricks to perform an efficient brute force integration are described in the supplementary material.  A key feature of the model is that for given values of $N_f$, $N_d$, $f_i$'s and $\sigma_j$, the model is linear.  Thus, the integration over the remaining parameters can be performed via linear algebra and the Laplace approximation.  Marginalizing over $\sigma_j$ and the $f_i$'s can be performed using standard numerical techniques.  
By evaluating the model conditioned on $N_f$ and $N_d$ and marginalizing over the remaining parameters, one can compare the marginal posterior probabilities to identify values of $N_f$ and $N_d$ that provide a good model for the observations without introducing more model parameters than are justified given the available data.  For our typical applications, the posterior dominated by small values of $N_f$ and $N_d$, so higher values need not be considered explicitly.  Since the surrogate model can be integrated numerically, it provides a quantitative basis for Bayesian model comparison and/or model selection.  

\section{Discussion of Application to Exoplanet Observations}

\subsection{Doppler Data}

One common challenge for exoplanet searches is deciding when the available data provides sufficient evidence to constitute a planet discovery.  This is particularly challenging in the case of stars with multiple planets, as the data can always be better modeled by adding additional planets.  The Fourier decomposition of the Doppler signature of a planet is dominated by power at the frequency ($1/P$) corresponding to the orbital period ($P$).  The power at the harmonic frequencies, $2/P$, $3/P$, ..., $\lambda/P$, decreases as $e^{\lambda-1}$, where $e$ is the orbital eccentricity.  The eccentricity for an ellipse is constrained to $(0,1)$ and most known exoplanets have eccentricities smaller than $0.15$.  Thus, applying the Bayesian surrogate model to Doppler data sets is expected to result in a posterior distribution for the most significant frequency ($f_1$) near one over the orbital period of the planet which dominates the Doppler signature.  The second most prominent frequency ($f_2$) could correspond to a harmonic of the first planet or to the orbital period of a second planet.  Some planetary systems contain two planets with orbital periods that differ by a factor of nearly 2 \citep[e.g.,][]{Laughlin05,Wright11,Lissauer11b}.  For certain planet masses and eccentricities, the Doppler signature of two such planets can be mimicked by one planet with a more eccentric orbit \citep{Giuppone09,Anglada10,Moorhead10}.  If there is actually only one planet, then one would expect the next most prominent frequency to correspond to twice the fundamental frequency ($f_2=2\times~f_1$) to within measurement precision.  On the other hand, for systems that actually contain two planets, the sidereal orbital periods often deviate from exact resonance \citep{Lissauer11b}.  
One might hope that in cases where the ratio of orbital periods differed from two, that the surrogate model could recognize this difference ($\delta\equiv~f_2-2\times~f_1$), so one could infer that the observations were due to two planets, rather than one planet on a more highly eccentric orbit \citep{Giuppone09,Anglada10,Moorhead10}.

To explore this possibility, we applied the Bayesian surrogate model to several sets of Doppler observations of exoplanet host stars.  We focused on exoplanets with a large velocity amplitudes and believed to have a significant eccentricity, as those systems provide the best prospects for measuring the harmonic frequencies precisely.  In particular, we choose systems with $K e^2 > 3$m/s, where $K$ is the velocity amplitude and $e$ is the orbital eccentricity.  In all cases, the surrogate model efficiently found the fundamental frequency corresponding to the orbital period.  As expected, $f_1$ is very tightly constrained and the marginalized posterior for additional frequencies are significantly broader.  In some cases, we found that the marginalized posterior for $f_2$ did not correspond to $2\times~f_1$.  

\begin{figure}[ht]
\begin{center}
\epsfig{file=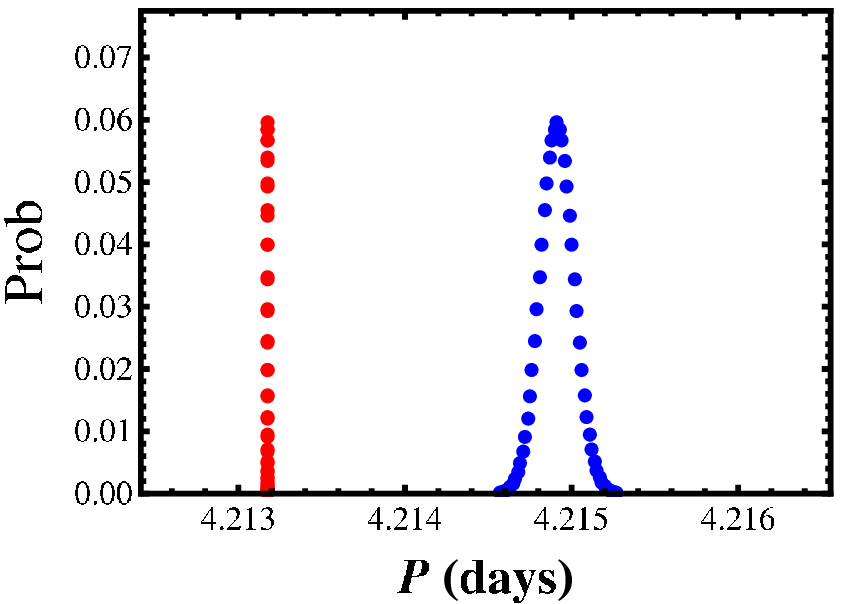}
\end{center}
\caption{In this figure, we consider a simulated data set based on the best-fit orbital parameters for the exoplanet HD 162020.  We show the marginalized posterior probability distributions for $P_1/2=2/f_1$ (red; very narrow distribution) and $P_2=1/f_2$ (blue; broad distribution) from a surrogate model with $N_f=2$.  Based on the Fourier expansion of the Doppler signature of a planet on a Keplerian orbit, one would expect that the surrogate model would yield marginalized posterior probability distributions for $f_2$ which overlaps marginalized posterior probability for $f_1/2$.  We find that this is not necessarily the case, even for this simulated data set with high signal-to-noise, uncorrelated Gaussian measurement errors and accurate estimates of the measurement uncertainties.  For this calculation, we have used the actual times of observations.  }
\label{fig_rv_act}
\end{figure}

\begin{figure}[ht]
\begin{center}
\epsfig{file=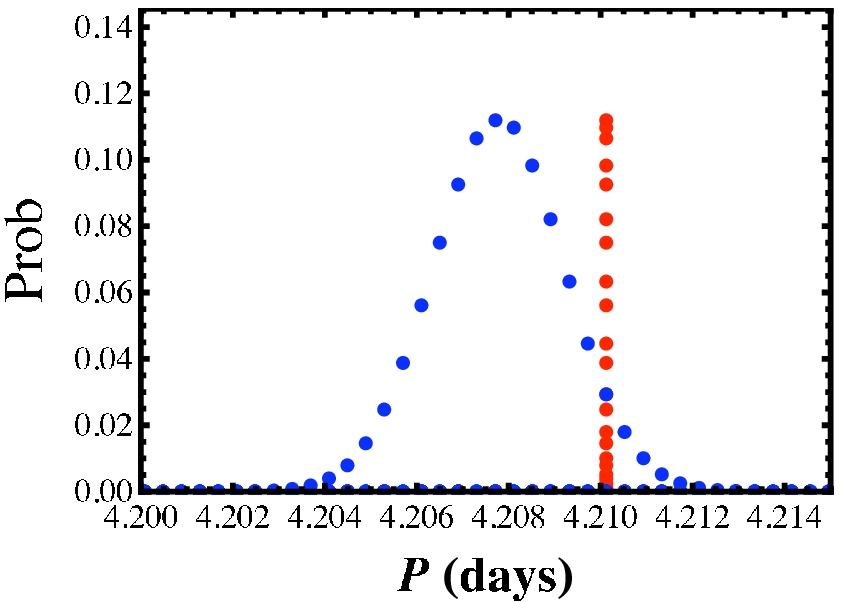}
\end{center}
\caption{This figure is analogous to Figure \ref{fig_rv_act}, except that the observation times are chosen randomly.  In this case, the posterior for $P_2=1/f_2$ (blue; broad distribution) overlaps with the posterior for $P_1/2=2f_1$ (red; narrow distribution), as expected for a Keplerian orbit.  Contrast this with Figure \ref{fig_rv_act} which uses actual observation times.  We conclude that for realistic Doppler data sets, aliasing due to limited number of the unevenly spaced observations can result in the marginal posterior distributions for $f_1/2$ and $f_2$ not overlapping.  This poses a considerable challenge to testing the one-planet model based on the dominant frequencies. }
\label{fig_rv_rand}
\end{figure}

In order to determine how often $f_2$ deviated from $2\times~f_1$ by chance, we performed a similar analysis on several simulated data sets.  Each simulation was modeled on  simulated velocities that were calculated according to the best-fit orbital period, amplitude, eccentricity, arrangement of pericenter and orbital phase.  We added Gaussian measurement noise with a scale set by the claimed measurement uncertainty.  The results for one case (HD 162020) are shown in Figures \ref{fig_rv_act} \& \ref{fig_rv_rand}.  
If we use the same observation times and uncertainties as the actual observations, then the marginal posterior distributions for $P_2=1/f_2$ and $P_1/2=2/f_1$ do not overlap.  If we generate a similar data set, but with random observation times, then the marginal posterior distributions for $P_2=1/f_2$ and $P_1/2=2/f_1$ do overlap.  We conclude that one must be very cautious of aliasing due to unevenly spaced observations when interpreting the marginal posterior distributions for $f_i$'s for realistic data sets.  

Our results suggest that the problems of aliasing could be reduced by obtaining regularly (or randomly) spaced observations.  Unfortunately, this is impractical for observations made from the surface of the Earth.  First, observations can not be made when the Sun is above the horizon (or even less than $\sim~12^\circ$ below the horizon) due to scattering of sunlight by Earth's atmosphere.  As the Earth revolves around the Sun, the time of day at which a given star can be observed from a given site changes.  For most stars, there are multiple months each year when high-precision Doppler observations are not possible, since the star appears too close to the sun (after projecting onto the sky).  Thus, for most stars, Doppler observations are prone to aliasing at frequencies associated with the solar day and the solar year.  (In principle, observations of stars near the North or South pole from an observatory in the Arctic or Antarctic could avoid aliasing near the day.  However, there are no observatories with high-precision Doppler capability near either pole due to logistical issues.)  Second, while planet host stars are faint compared to the daytime sky brightness, they are much brighter than distant galaxies.  Therefore, time allocation committees assign the vast majority of observing time near new Moon to extragalactic astronomers.  Exoplanet searches are typically are assigned observing time near full Moon, introducing aliasing at frequencies associated with the lunar month ($\simeq~29.5$ days).  In practice, the best way to avoid this is to dedicate an observatory to Doppler observations (e.g., the HARPS instrument at the European Southern Observatory's 3.6m telescope in La Silla, Chile).  Of course, this requires considerable resources and is not an option for the world's largest telescopes.  Third, astronomers often attempt to optimize the efficiency of their observations on a given night by observing each star when it is near the greatest (angular) altitude in the sky that night, as this minimizes the amount of absorption of starlight by the Earth's atmosphere.  This strategy introduces yet another frequency associated with the sidereal day (i.e., the time for a star to return to the nearly same point on the sky from a given location, roughly 23 hours, 56 minutes and 4 seconds).  While these may seem like picky details, each of these timescales can be found in public data and in some cases contribute to qualitative ambiguities in the orbital solutions \citep{DawsonFab10}.  While attention to scheduling can help improve efficiency of planet searches, ultimately weather (i.e., cloudy skies) will lead to data gaps and prevent optimal experimental design for any Earth-based observatory.  Space-based observatories are orders of magnitude more expensive to construct and operate.  Therefore, we must develop tools to analyze realistic data sets.  Studies such as this will help us to use those tools responsibly and reduce the risk of making erroneous claims.

\subsection{Transit Timing Variations}

The rapidly increasing number of known exoplanets that transit their host star has led several observers to measure transit times in hopes of detecting deviations from a linear ephemeris due to perturbations by another (potentially non-transiting) planet \citep[e.g.,][]{MillerRicci08,Maciejewski10}.  In most previous studies, the number of precisely measured transit times has been insufficient to fit a physical model.  The prospects for the transit timing variation method are poised to improve dramatically thanks to NASA's {\em Kepler} mission, which is observing over 100,000 stars nearly continuously for 3.5 years \citep{Borucki11,Steffen10}.  Indeed, {\em Kepler} recently reported the first strong evidence for transiting timing variations in Kepler-9 and Kepler-11 systems \citep{Holman10,Lissauer11a}.  Interestingly, in both of these cases, transit timing variations were used to confirm planet candidates that had already been identified via transit and to constrain their masses and orbits.  {\em Kepler} has also identified dozens of planet candidates with putative transit timing variations \citep{Ford11}.  However, further observations are needed in order to infer the mass and orbital properties of the perturbing bodies.  

In principle, one could model the full light curve \citep[i.e., observed brightness versus time;][]{Carter11}.  However, this would vastly increase the computational time required and most of the information is contained in the transit time. (Using data from NASA's {\em Kepler} mission, there are typically $\sim~10^3-10^7$ brightness measurements for each transit time.)  The transit time is the most precisely measured parameter for each transit, and the transit times are sensitive to whether the transiting planet is slightly ahead or behind ``schedule'' due to gravitational perturbations from other planets.  The next best-measured parameter is the transit duration which depend on the orbit of the transiting planet and the stellar radius \citep{Moorhead11}.  The depth of each transit is primarily determined by the relative sizes of the planet and host star and is not affected by gravitational perturbations of other planets.  The detailed shape of each transit also depends on detailed stellar properties (``limb darkening parameters'').  For a system of non-coplanar planets, transit duration variations may be detectable.  We have focused our analysis in this paper on coplanar systems viewed edge-on, so as to reduce the dimensionality of the parameter space to be explored.  While more parameters are required to describe non-coplanar systems, in some cases it may be possible to derive additional constraints on the orbits based on transit duration variations, or lack thereof \citep{Holman10}.  Given the computational cost of modeling the full light curve, we recommend future research to develop tools to analyze a series of transit times and transit durations.

To explore the potential for transit timing variations to enable the detection of a non-transiting planet, we have generated $\sim 10^7$ simulated data sets with a wide variety of orbital periods, eccentricities and angles \citep{Veras11}.  We apply the surrogate model to simulated data sets to identify the dominant frequency and its amplitude.  We find that the surrogate model can provide an accurate model for some data sets, particularly those very near a mean motion resonance, a regime which is particularly difficult to approximate analytically \citep{Nesvorny08,Nesvorny09,Nesvorny10}.  In other cases, the transit timing variation signature is more complex and would require several frequencies to model adequately.  For extended data sets, this can be challenging, both due to computation time and available memory.  

While the surrogate model can provide a reasonable approximation to many transit timing signatures, the inferred model parameters depend sensitively on the orbital phases, as well as more basic physical parameters such as the planet mass and orbital period.  Further, as one increases the number of observations, the inferred parameters often change significantly.  This significantly complicates the interpretation of the surrogate model outputs.  While the inferred model parameters will eventually stabilize with a sufficient number and time span of observations, we find that inferred parameters can continue to change even after several years of observations.  We conclude that it may not always be practical to invert transit timing variations and infer the mass and orbital properties of a non-transiting planet \citep{Ragozzine10}.  A more extended discussion of implications for transit timing planet searches is presented separately \citep{Payne10,Veras11}.  Of course, our results do not prove that other analysis techniques can not invert transit timing variations.  However, our surrogate model was designed to capture the most important aspects of the problem.  Thus, our results are suggestive that this problem may be more general.  Since the original submission of this paper, the first confirmations of exoplanets via the transit timing variation methods were published \citep{Holman10,Lissauer11a}.  In these cases, each of the detected planets transit the star, so transit timing variations were used to confirm planet candidates that had already been identified by the standard transit technique.  \cite{Ford11} showed that {\em Kepler} can be expected to measure transit timing variations for at least 12 systems with multiple transiting planet candidates.  Based on analysis of the frequency of multiple transiting planet candidate systems \citep{Lissauer11b}, we expect that even more planets with transit timing variations will be significantly perturbed by a non-transiting planet.  Indeed, based on the first four months of observations, {\em Kepler} has identified dozens of planet candidates with prospective transit timing variations, most in systems with only a single transiting planet candidate.  Both our results and \cite{Ford11} suggest that further observations will be necessary before the masses and orbits of putative additional planets can be determined.  

\subsection{Conclusions}
We developed a Bayesian surrogate model for analysis of time series data in general and applied this model to two types of exoplanet search data, Doppler and transit timing variations.  The surrogate model can be evaluated very rapidly and we describe a method for efficiently integrating over most model parameters.  This allows for calculating the (properly normalized) marginalized posterior probability and assessing the posterior probability for a given periodicity.  

One strength of the surrogate model is for exploratory data analysis.  For example, astronomers routinely use the Lomb-Scargle periodogram to search Doppler data for a periodic signature of a planet and to identify the range of periods that should be explored with a more detailed model.  One limiting case of the surrogate model ($N_{f,max}=1$ and $N_{d,max}=0$) directly corresponds to the Bayesian generalization of the Lomb-Scargle periodogram \citep{Cumming04}.
When a planet has a large eccentricity or one star hosts multiple planets, the  Lomb-Scargle periodogram typically reveals multiple significant periodicities.  Previously, astronomers have dealt with this by applying the Lomb-Scargle periodogram to the residuals after subtracting the best-fit sinusoidal or Keplerian model.  This approach can bias subsequent results, since the subtracted model is not exact.  Further, assessing the significance of peaks in the periodogram of residuals is nontrivial.  Most authors use a blind approach when searching for additional periodicities, but others favor using information about the frequencies previously identified \citep[e.g.,][]{Konacki99,DawsonFab10}.  In practice, this can lead to a combersome decission tree in a frequentist context.  Our surrogate model (with $N_{f,max}>1$) represents a Bayesian generalization of iterative frequentist methods for analyzing periodogram of residuals.  The Bayesian surrogate model provides a rigorous basis for calculating Bayes factor of the marginalized posterior probability for the number of significant frequencies $B_{n+1,n}=p(N_f=n+1|x,y,\sigma)/p(N_f=n|x,y,\sigma)$.  Another advantage of the surrogate model is that by marginalizing over the other model parameters (e.g., frequencies, amplitudes, jitter, polynomial terms), a spurious false positive should be less likely than  analyzing residuals to only the best-fit model.  

Finally, for many systems the surrogate model can provide a lower-dimensional model that still captures the important (i.e., observable) physical effects.  For example, in a system of multiple low-mass planets, a full physical model has a dimension of $\simeq~7N_p$, where $N_p$ is the number of planets.  If the system has planetary
orbits with small eccentricities and/or inclinations, then several of the model parameters may have no observable effect.  For such systems, the surrogate model would be able to describe the system accurately using a lower-dimensional parameter space ($\simeq~3N_p$), greatly increasing computational efficiency and perhaps increasing the sensitivity for detecting additional planets (due to the less extreme Occam's factor).  

The surrogate model is not meant to replace other tools for Bayesian parameter estimation and model selection.  It is still beneficial to apply MCMC (and variants) for parameter estimation using a more physical model \citep[e.g.,][]{Ford05,Gregory05,Ford06,JohnsonPayne11}.  Similarly, tools such as restricted Monte Carlo, importance sampling and parallel tempering can be helpful for calculating Bayes factors using a more physical model \citep[e.g.,][]{FordGregory07,Gregory11}.  Since these tools are much more computationally expensive than the surrogate model, they are most appropriate once a putative set of orbital periods has been identified (e.g., by periodogram analysis, surrogate model, or human inspection for sufficiently large signals).  

Thanks to the computational efficiency of the surrogate model, we were able to analyze numerous simulated datasets including multiple planet systems, something that would not have been feasible using previously available Bayesian methods such as MCMC.  For Doppler planet searches we find that realistic observing cadences can lead to significant aliasing that prevents precisely testing whether there is a harmonic relationship between measured frequencies.  Given the close relationship of our surrogate model to the widely used Lomb-Scargle periodogram, our results also serve as caution regarding the integration of results based on periodogram analyses \citep[e.g.,][]{Konacki99,Anglada10,DawsonFab10}.  

For analyzing transit timing variations, we find the posterior distribution for the surrogate model parameters are sensitive to the exact orbital configuration.  While the sensitivity to important physical parameters is advantageous, sensitivity to parameters that do not have dynamical significance makes interpretation of the posterior distribution for surrogate model parameters more challenging. Unfortunately, the transit timing signature often evolves on a timescale comparable to or longer than a realistic time spans for observations (e.g., 3.5-10 years for {\em Kepler}).  This makes it impractical to build a library of possible transit timing signatures and the corresponding surrogate model outputs.  The surrogate model may still be useful for establishing the significance of putative periodicities and/or long-term trends in transit timing data.  In addition to the advantages of a Bayesian approach, the computational efficiency of the surrogate model could be useful for analyzing large sets of simulated data sets to aid in interpretation of a transit timing variation planet search. 

In both cases, we find that qualitative results (e.g., whether the Bayes factor, $B_{n,n+1}$, for the significance of additional frequency is greater or less than unity) can depend on the choice of prior for the jitter parameter ($\sigma_j$).  In this paper we used a mathematically motivated prior for $\sigma_j$.  Our result suggests that practical application of the surrogate model would significantly benefit from further astronomical observations and statistical analyses to assess the empirical distribution of the stellar jitter for both Doppler \citep[e.g.,][]{Wright05} and transit timing observations \citep[e.g.,][]{Holman10}.  Given the relationship of our surrogate model to non-Bayesian methods being employed by astronomers, our results also suggest caution in the interpretation of other results based on frequentist analyses (which typically assume a single fixed value of the jitter).

\section{Supplementary Material}

\subsection{Practical Model Evaluation}
\subsubsection{Integration over Linear Parameters}
The surrogate model is linear in the parameters: $S_i$'s, $C_i$'s, and $D_i$'s.  For a given choice of $N_f$, $N_d$, $f_i$'s and $\sigma_j$, one can calculate the ``best-fit'' values of $S_i$'s, $C_i$'s, and $D_i$'s via simple linear algebra.  The same linear algebraic operations allow the integrals over these linear parameters to be evaluated efficiently via the Laplace approximation (i.e., we expand the exponent about the best-fit parameters, keep the constant and second order terms, and extend the limits of integration to infinity; \citet{Cumming04}).  For our applications, the posterior probability is typically a smooth function of $\sigma_j$'s, but can vary extremely rapidly with $f_i$'s.  Thus, we recommend fixing the choice of $N_f$, $N_d$, and $f_i$'s, as that the integral over $\sigma_j$ can be evaluated efficiently with standard numerical integration
 techniques.  (In the special case of equal measurement uncertainties, the integrals can be calculated analytically.)   

\subsubsection{Brute Force Integration over Frequencies}
Unfortunately, the integrals over the $f_i$'s must be evaluated numerically.  We recommend discretizing these integrals and evaluating them via brute force.  The number of frequencies to be evaluated ($M$) can be quite large   \citep{Cumming04}.  Fortunately, there are several computational tricks that can speed up the calculation.  In particular, most of the trigonometric functions can be computed using trigonometric identities to improve performance.  We find that brute force integration over one or even two $f_i$'s is practical for realistic data sets.  (For large data sets, a large amount of RAM may be required for efficient evaluation.) 

\subsubsection{Approximations for Models with Many Frequencies}
Unfortunately, brute force integration over more than two $f_i$'s rapidly become prohibitive.  Thus, we introduce the following approximation.  We start by performing a brute force evaluation of the model conditioned on there being one frequency and then successively approximate the posterior conditioned on two frequencies, three frequencies, etc..  When approximating the model conditioned on there being $N_f$ frequencies (and $N_f\ge~2$), we limit the set of $f_i$'s with $i<N_f$ that are considered to those which contributed significantly to the posterior probability for the model conditioned on there being $N_f-1$ frequencies.  That is we set a threshold (e.g., $\epsilon=10^{-4}$) and after evaluating the model conditioned on $N_f=1$ frequencies, we store the $m_{1}$ frequencies which have the largest posterior probabilities and collectively sum to at least $1-\epsilon$ of the posterior probability conditioned on there being 1 frequency.  At this point, we have searched $M\times~(1+m_1)$ frequencies (rather than $M^2$).  Next, we estimate the marginal posterior probability for $N_f=3$ by considering only those combinations of $f_1$ and $f_2$ that have the largest posterior probabilities and collectively sum to at least $1-\epsilon$ of the posterior probability conditioned on there being 2 frequencies.  Thus, when evaluating the model conditioned on $N_f$ frequencies, there are no more than $M \prod_{i=1}^{N_f-1} m_{i}$ frequencies to be explicitly evaluated, much less than $M^{N_f}$.  In principle, this approximation can be relaxed by performing a full search over two frequencies.  In this case evaluating the model conditioned on $N_f$ frequencies, requires  no more than $M^2 \prod_{i=1}^{N_f-2} m_{i}$ frequencies to be explicitly evaluated, which may or may not be practical for a given data set.  

\subsubsection{Truncating the Number of Frequencies}

For data sets which are well described be only one or a few frequencies, models which include a large number of frequencies will have a very small posterior probability due to the Occam's factor associated with the higher-dimensional model.  Thus, there is little point in evaluating models with large $N_f$. Thus, we recommend successively calculating the posterior probability conditioned on $N_f$ ($p(\theta|N_f,x_k,y_k,\sigma_k)$) and stopping at $N_{f,stop}$, such that
$p(N_f=N_{f,stop}|x_k,y_k,\sigma_k) \ll \sum_{i=0}^{N_{f,stop}-1} p(N_f=i|x_k,y_k,\sigma_k)$ \\
and approximate the remaining models as $p(N_f>N_{f,stop}|x_k,y_k,\sigma_k)\approx 0$.

\subsubsection{Computational Cost}
The surrogate model can be evaluated much more quickly than an n-body integration, explores a lower-dimensional parameter space, and takes advantage of the linearity of the model in most of the model parameters.  Nevertheless, the strong sensitivity to frequency dictates that we must perform a fine sampling in frequency.  
For example, consider a case of a $\simeq~10 M_{Earth}$-mass transiting planet with an orbital period near 4 days and a small planet which is not observed to transit with a period near 8 days.  With a modest eccentricity (0.1), the transit timing variations of the inner planet could be $\sim~10$ minutes, comparable to the timing precision for each transit for a typical {\em Kepler} planet host star \citep{Ford11}.  Over 7.5 years of observations (possible with an extended {\em Kepler} mission), one would observe roughly 680 transits, allowing for an easy detection of such a single.  If we assume a timing precision of 10 minutes and set $f_{\max}$ to 1 days, then each integral over frequency requires considering $M\sim~180,000$ frequencies.  With a single core of a AMD Opteron 275 processor (2.2 GHz), this takes $\sim~10$, 16 or 37 seconds for models with $N_f=1$, 2 or 3.  Setting $\epsilon=10^{-3}$ (see supplementary materials), we needed to compute 1 ($N_f=1$), 11 ($N_f=2$) and 16 ($N_f=3$) scans over frequency for both $N_d=0$ and 1, requiring a total of 12 CPU-minutes for one system.  Thus, the brute force exploration is relatively fast for a single model.  For systems with no detectable signal, the time required decreases significantly, as $N_f=3$ (or even 2) need not be explored.  On the other hand, the computation time per system required grows significantly as the signal-to-noise increases, since the number of frequencies sampled ($M$) must be increased to avoid missing a narrow peak in the posterior density.  Of course, in these cases, the signal is sufficiently large that fancy statistical methods are not necessary to detect the dominant signal.  The speed of the surrogate model allowed us to analyze millions of simulated data sets and to explore the complex parameter space, using a cluster with hundreds of AMD Opteron servers at the University of Florida High-Performance Computing Center.

\bibliographystyle{ba}
\bibliography{main}{}

\begin{thebibliography}{42}
\newcommand{\enquote}[1]{``#1''}
\expandafter\ifx\csname natexlab\endcsname\relax\def\natexlab#1{#1}\fi
\expandafter\ifx\csname url\endcsname\relax
  \def\url#1{{\tt #1}}\fi
\expandafter\ifx\csname urlprefix\endcsname\relax\def\urlprefix{URL }\fi

\bibitem[{{Agol} et~al.(2005){Agol}, {Steffen}, {Sari}, and
  {Clarkson}}]{Agol05}
{Agol}, E., {Steffen}, J., {Sari}, R., and {Clarkson}, W. (2005).
\newblock \enquote{{On detecting terrestrial planets with timing of giant
  planet transits}.}
\newblock {\em \mnras\/}, 359: 567--579.

\bibitem[{{Anglada-Escud{\'e}} et~al.(2010){Anglada-Escud{\'e}},
  {L{\'o}pez-Morales}, and {Chambers}}]{Anglada10}
{Anglada-Escud{\'e}}, G., {L{\'o}pez-Morales}, M., and {Chambers}, J.~E.
  (2010).
\newblock \enquote{{How Eccentric Orbital Solutions Can Hide Planetary Systems
  in 2:1 Resonant Orbits}.}
\newblock {\em \apj\/}, 709: 168--178.

\bibitem[{{Borucki} et~al.(2011){Borucki}, {Koch}, {Basri}, {Batalha}, {Brown},
  {Bryson}, {Caldwell}, {Christensen-Dalsgaard}, {Cochran}, {DeVore}, {Dunham},
  {Gautier}, {Geary}, {Gilliland}, {Gould}, {Howell}, {Jenkins}, {Latham},
  {Lissauer}, {Marcy}, {Rowe}, {Sasselov}, {Boss}, {Charbonneau}, {Ciardi},
  {Doyle}, {Dupree}, {Ford}, {Fortney}, {Holman}, {Seager}, {Steffen},
  {Tarter}, {Welsh}, {Allen}, {Buchhave}, {Christiansen}, {Clarke},
  {D{\'e}sert}, {Endl}, {Fabrycky}, {Fressin}, {Haas}, {Horch}, {Howard},
  {Isaacson}, {Kjeldsen}, {Kolodziejczak}, {Kulesa}, {Li}, {Machalek},
  {McCarthy}, {MacQueen}, {Meibom}, {Miquel}, {Prsa}, {Quinn}, {Quintana},
  {Ragozzine}, {Sherry}, {Shporer}, {Tenenbaum}, {Torres}, {Twicken}, {Van
  Cleve}, and {Walkowicz}}]{Borucki11}
{Borucki}, W.~J., {Koch}, D.~G., {Basri}, G., {Batalha}, N., {Brown}, T.~M.,
  {Bryson}, S.~T., {Caldwell}, D., {Christensen-Dalsgaard}, J., {Cochran},
  W.~D., {DeVore}, E., {Dunham}, E.~W., {Gautier}, T.~N., III, {Geary}, J.~C.,
  {Gilliland}, R., {Gould}, A., {Howell}, S.~B., {Jenkins}, J.~M., {Latham},
  D.~W., {Lissauer}, J.~J., {Marcy}, G.~W., {Rowe}, J., {Sasselov}, D., {Boss},
  A., {Charbonneau}, D., {Ciardi}, D., {Doyle}, L., {Dupree}, A.~K., {Ford},
  E.~B., {Fortney}, J., {Holman}, M.~J., {Seager}, S., {Steffen}, J.~H.,
  {Tarter}, J., {Welsh}, W.~F., {Allen}, C., {Buchhave}, L.~A., {Christiansen},
  J.~L., {Clarke}, B.~D., {D{\'e}sert}, J., {Endl}, M., {Fabrycky}, D.,
  {Fressin}, F., {Haas}, M., {Horch}, E., {Howard}, A., {Isaacson}, H.,
  {Kjeldsen}, H., {Kolodziejczak}, J., {Kulesa}, C., {Li}, J., {Machalek}, P.,
  {McCarthy}, D., {MacQueen}, P., {Meibom}, S., {Miquel}, T., {Prsa}, A.,
  {Quinn}, S.~N., {Quintana}, E.~V., {Ragozzine}, D., {Sherry}, W., {Shporer},
  A., {Tenenbaum}, P., {Torres}, G., {Twicken}, J.~D., {Van Cleve}, J., and
  {Walkowicz}, L. (2011).
\newblock \enquote{{Characteristics of planetary candidates observed by Kepler,
  II: Analysis of the first four months of data}.}
\newblock {\em \apj\/}, 736: 19--40.

\bibitem[{{Carter} et~al.(2011){Carter}, {Fabrycky}, {Ragozzine}, {Holman},
  {Quinn}, {Latham}, {Buchhave}, {Van Cleve}, {Cochran}, {Cote}, {Endl},
  {Ford}, {Haas}, {Jenkins}, {Koch}, {Li}, {Lissauer}, {MacQueen}, {Middour},
  {Orosz}, {Rowe}, {Steffen}, and {Welsh}}]{Carter11}
{Carter}, J.~A., {Fabrycky}, D.~C., {Ragozzine}, D., {Holman}, M.~J., {Quinn},
  S.~N., {Latham}, D.~W., {Buchhave}, L.~A., {Van Cleve}, J., {Cochran}, W.~D.,
  {Cote}, M.~T., {Endl}, M., {Ford}, E.~B., {Haas}, M.~R., {Jenkins}, J.~M.,
  {Koch}, D.~G., {Li}, J., {Lissauer}, J.~J., {MacQueen}, P.~J., {Middour},
  C.~K., {Orosz}, J.~A., {Rowe}, J.~F., {Steffen}, J.~H., and {Welsh}, W.~F.
  (2011).
\newblock \enquote{{KOI-126: A Triply Eclipsing Hierarchical Triple with Two
  Low-Mass Stars}.}
\newblock {\em Science\/}, 331: 562--.

\bibitem[{{Cumming}(2004)}]{Cumming04}
{Cumming}, A. (2004).
\newblock \enquote{{Detectability of extrasolar planets in radial velocity
  surveys}.}
\newblock {\em \mnras\/}, 354: 1165--1176.

\bibitem[{{Cumming} et~al.(2008){Cumming}, {Butler}, {Marcy}, {Vogt}, {Wright},
  and {Fischer}}]{Cumming08}
{Cumming}, A., {Butler}, R.~P., {Marcy}, G.~W., {Vogt}, S.~S., {Wright}, J.~T.,
  and {Fischer}, D.~A. (2008).
\newblock \enquote{{The Keck Planet Search: Detectability and the Minimum Mass
  and Orbital Period Distribution of Extrasolar Planets}.}
\newblock {\em \pasp\/}, 120: 531--554.

\bibitem[{{Dawson} and {Fabrycky}(2010)}]{DawsonFab10}
{Dawson}, R.~I. and {Fabrycky}, D.~C. (2010).
\newblock \enquote{{Radial Velocity Planets De-aliased: A New, Short Period for
  Super-Earth 55 Cnc e}.}
\newblock {\em \apj\/}, 722: 937--953.

\bibitem[{{Ford}(2005)}]{Ford05}
{Ford}, E.~B. (2005).
\newblock \enquote{{Quantifying the Uncertainty in the Orbits of Extrasolar
  Planets}.}
\newblock {\em \aj\/}, 129: 1706--1717.

\bibitem[{{Ford}(2006)}]{Ford06}
--- (2006).
\newblock \enquote{{Improving the Efficiency of Markov Chain Monte Carlo for
  Analyzing the Orbits of Extrasolar Planets}.}
\newblock {\em \apj\/}, 642: 505--522.

\bibitem[{{Ford}(2008)}]{Ford08}
--- (2008).
\newblock \enquote{{Adaptive Scheduling Algorithms for Planet Searches}.}
\newblock {\em \aj\/}, 135: 1008--1020.

\bibitem[{{Ford} and {Gregory}(2007)}]{FordGregory07}
{Ford}, E.~B. and {Gregory}, P.~C. (2007).
\newblock \enquote{{Bayesian Model Selection and Extrasolar Planet Detection}.}
\newblock In {G.~J.~Babu \& E.~D.~Feigelson} (ed.), {\em Statistical Challenges
  in Modern Astronomy IV\/}, volume 371 of {\em Astronomical Society of the
  Pacific Conference Series\/}, 189--+.

\bibitem[{{Ford} and {Holman}(2007)}]{FordHolman07}
{Ford}, E.~B. and {Holman}, M.~J. (2007).
\newblock \enquote{{Using Transit Timing Observations to Search for Trojans of
  Transiting Extrasolar Planets}.}
\newblock {\em \apjl\/}, 664: L51--L54.

\bibitem[{{Ford} et~al.(2011){Ford}, {Rowe}, {Fabrycky}, {Carter}, {Holman},
  {Lissauer}, {Ragozzine}, {Steffen}, {Borucki}, {Batalha}, {Bryson},
  {Caldwell}, {Gautier}, {Jenkins}, {Koch}, {Li}, {Lucas}, {Marcy}, {Mullally},
  {Quintana}, {McCauliff}, {Thompson}, {Tenenbaum}, {Still}, and
  {Twicken}}]{Ford11}
{Ford}, E.~B., {Rowe}, J.~F., {Fabrycky}, D.~C., {Carter}, J., {Holman}, M.~J.,
  {Lissauer}, J.~J., {Ragozzine}, D., {Steffen}, J.~H., {Borucki}, W.~J.,
  {Batalha}, N.~M., {Bryson}, S., {Caldwell}, D.~A., {Gautier}, T.~N., III,
  {Jenkins}, J.~M., {Koch}, D.~G., {Li}, J., {Lucas}, P., {Marcy}, G.~W.,
  {Mullally}, F.~R., {Quintana}, E., {McCauliff}, S., {Thompson}, S.~E.,
  {Tenenbaum}, P., {Still}, M., and {Twicken}, J.~D. (2011).
\newblock \enquote{{Transit Timing Observations from Kepler: I. Statistical
  Analysis of the First Four Months}.}
\newblock {\em \apjs\/}.

\bibitem[{{Giuppone} et~al.(2009){Giuppone}, {Tadeu dos Santos}, {Beaug{\'e}},
  {Ferraz-Mello}, and {Michtchenko}}]{Giuppone09}
{Giuppone}, C.~A., {Tadeu dos Santos}, M., {Beaug{\'e}}, C., {Ferraz-Mello},
  S., and {Michtchenko}, T.~A. (2009).
\newblock \enquote{{Detectability and Error Estimation in Orbital Fits of
  Resonant Extrasolar Planets}.}
\newblock {\em \apj\/}, 699: 1321--1332.

\bibitem[{{Gregory}(2005)}]{Gregory05}
{Gregory}, P.~C. (2005).
\newblock \enquote{{A Bayesian Analysis of Extrasolar Planet Data for HD
  73526}.}
\newblock {\em \apj\/}, 631: 1198--1214.

\bibitem[{{Gregory}(2011)}]{Gregory11}
--- (2011).
\newblock \enquote{{Bayesian Re-analysis of the Gliese 581 Exoplanet System}.}
\newblock {\em \mnras\/}.

\bibitem[{{Holman} et~al.(2010){Holman}, {Fabrycky}, {Ragozzine}, {Ford},
  {Steffen}, {Welsh}, {Lissauer}, {Latham}, {Marcy}, {Walkowicz}, {Batalha},
  {Jenkins}, {Rowe}, {Cochran}, {Fressin}, {Torres}, {Buchhave}, {Sasselov},
  {Borucki}, {Koch}, {Basri}, {Brown}, {Caldwell}, {Charbonneau}, {Dunham},
  {Gautier}, {Geary}, {Gilliland}, {Haas}, {Howell}, {Ciardi}, {Endl},
  {Fischer}, {F{\"u}r{\'e}sz}, {Hartman}, {Isaacson}, {Johnson}, {MacQueen},
  {Moorhead}, {Morehead}, and {Orosz}}]{Holman10}
{Holman}, M.~J., {Fabrycky}, D.~C., {Ragozzine}, D., {Ford}, E.~B., {Steffen},
  J.~H., {Welsh}, W.~F., {Lissauer}, J.~J., {Latham}, D.~W., {Marcy}, G.~W.,
  {Walkowicz}, L.~M., {Batalha}, N.~M., {Jenkins}, J.~M., {Rowe}, J.~F.,
  {Cochran}, W.~D., {Fressin}, F., {Torres}, G., {Buchhave}, L.~A., {Sasselov},
  D.~D., {Borucki}, W.~J., {Koch}, D.~G., {Basri}, G., {Brown}, T.~M.,
  {Caldwell}, D.~A., {Charbonneau}, D., {Dunham}, E.~W., {Gautier}, T.~N.,
  {Geary}, J.~C., {Gilliland}, R.~L., {Haas}, M.~R., {Howell}, S.~B., {Ciardi},
  D.~R., {Endl}, M., {Fischer}, D., {F{\"u}r{\'e}sz}, G., {Hartman}, J.~D.,
  {Isaacson}, H., {Johnson}, J.~A., {MacQueen}, P.~J., {Moorhead}, A.~V.,
  {Morehead}, R.~C., and {Orosz}, J.~A. (2010).
\newblock \enquote{{Kepler-9: A System of Multiple Planets Transiting a
  Sun-Like Star, Confirmed by Timing Variations}.}
\newblock {\em Science\/}, 330: 51--.

\bibitem[{{Holman} and {Murray}(2005)}]{HolmanMurray05}
{Holman}, M.~J. and {Murray}, N.~W. (2005).
\newblock \enquote{{The Use of Transit Timing to Detect Terrestrial-Mass
  Extrasolar Planets}.}
\newblock {\em Science\/}, 307: 1288--1291.

\bibitem[{{Howard} et~al.(2010){Howard}, {Marcy}, {Johnson}, {Fischer},
  {Wright}, {Isaacson}, {Valenti}, {Anderson}, {Lin}, and {Ida}}]{Howard10}
{Howard}, A.~W., {Marcy}, G.~W., {Johnson}, J.~A., {Fischer}, D.~A., {Wright},
  J.~T., {Isaacson}, H., {Valenti}, J.~A., {Anderson}, J., {Lin}, D.~N.~C., and
  {Ida}, S. (2010).
\newblock \enquote{{The Occurrence and Mass Distribution of Close-in
  Super-Earths, Neptunes, and Jupiters}.}
\newblock {\em Science\/}, 330: 653--.

\bibitem[{{Johnson} et~al.(2011){Johnson}, {Payne}, {Howard}, {Clubb}, {Ford},
  {Bowler}, {Henry}, {Fischer}, {Marcy}, {Brewer}, {Schwab}, {Reffert}, and
  {Lowe}}]{JohnsonPayne11}
{Johnson}, J.~A., {Payne}, M., {Howard}, A.~W., {Clubb}, K.~I., {Ford}, E.~B.,
  {Bowler}, B.~P., {Henry}, G.~W., {Fischer}, D.~A., {Marcy}, G.~W., {Brewer},
  J.~M., {Schwab}, C., {Reffert}, S., and {Lowe}, T.~B. (2011).
\newblock \enquote{{Retired a Stars and Their Companions. VI. A Pair of
  Interacting Exoplanet Pairs Around the Subgiants 24 Sextanis and HD 200964}.}
\newblock {\em \aj\/}, 141: 16--+.

\bibitem[{{Konacki} and {Maciejewski}(1999)}]{Konacki99}
{Konacki}, M. and {Maciejewski}, A.~J. (1999).
\newblock \enquote{{Frequency Analysis of Reflex Velocities of Stars with
  Planets}.}
\newblock {\em \apj\/}, 518: 442--449.

\bibitem[{{Laughlin} et~al.(2005){Laughlin}, {Butler}, {Fischer}, {Marcy},
  {Vogt}, and {Wolf}}]{Laughlin05}
{Laughlin}, G., {Butler}, R.~P., {Fischer}, D.~A., {Marcy}, G.~W., {Vogt},
  S.~S., and {Wolf}, A.~S. (2005).
\newblock \enquote{{The GJ 876 Planetary System: A Progress Report}.}
\newblock {\em \apj\/}, 622: 1182--1190.

\bibitem[{{Lissauer} et~al.(2011{\natexlab{a}}){Lissauer}, {Fabrycky}, {Ford},
  {Borucki}, {Fressin}, {Marcy}, {Orosz}, {Rowe}, {Torres}, {Welsh}, {Batalha},
  {Bryson}, {Buchhave}, {Caldwell}, {Carter}, {Charbonneau}, {Christiansen},
  {Cochran}, {Desert}, {Dunham}, {Fanelli}, {Fortney}, {Gautier}, {Geary},
  {Gilliland}, {Haas}, {Hall}, {Holman}, {Koch}, {Latham}, {Lopez},
  {McCauliff}, {Miller}, {Morehead}, {Quintana}, {Ragozzine}, {Sasselov},
  {Short}, and {Steffen}}]{Lissauer11a}
{Lissauer}, J.~J., {Fabrycky}, D.~C., {Ford}, E.~B., {Borucki}, W.~J.,
  {Fressin}, F., {Marcy}, G.~W., {Orosz}, J.~A., {Rowe}, J.~F., {Torres}, G.,
  {Welsh}, W.~F., {Batalha}, N.~M., {Bryson}, S.~T., {Buchhave}, L.~A.,
  {Caldwell}, D.~A., {Carter}, J.~A., {Charbonneau}, D., {Christiansen}, J.~L.,
  {Cochran}, W.~D., {Desert}, J., {Dunham}, E.~W., {Fanelli}, M.~N., {Fortney},
  J.~J., {Gautier}, T.~N., III, {Geary}, J.~C., {Gilliland}, R.~L., {Haas},
  M.~R., {Hall}, J.~R., {Holman}, M.~J., {Koch}, D.~G., {Latham}, D.~W.,
  {Lopez}, E., {McCauliff}, S., {Miller}, N., {Morehead}, R.~C., {Quintana},
  E.~V., {Ragozzine}, D., {Sasselov}, D., {Short}, D.~R., and {Steffen}, J.~H.
  (2011{\natexlab{a}}).
\newblock \enquote{{A closely packed system of low-mass, low-density planets
  transiting Kepler-11}.}
\newblock {\em \nat\/}, 470: 53--58.

\bibitem[{{Lissauer} et~al.(2011{\natexlab{b}}){Lissauer}, {Ragozzine},
  {Fabrycky}, {Steffen}, {Ford}, {Jenkins}, {Shporer}, {Holman}, {Rowe},
  {Quintana}, {Batalha}, {Borucki}, {Bryson}, {Caldwell}, {Carter}, {Ciardi},
  {Dunham}, {Fortney}, {Gautier}, {Howell}, {Koch}, {Latham}, {Marcy},
  {Morehead}, and {Sasselov}}]{Lissauer11b}
{Lissauer}, J.~J., {Ragozzine}, D., {Fabrycky}, D.~C., {Steffen}, J.~H.,
  {Ford}, E.~B., {Jenkins}, J.~M., {Shporer}, A., {Holman}, M.~J., {Rowe},
  J.~F., {Quintana}, E.~V., {Batalha}, N.~M., {Borucki}, W.~J., {Bryson},
  S.~T., {Caldwell}, D.~A., {Carter}, J.~A., {Ciardi}, D., {Dunham}, E.~W.,
  {Fortney}, J.~J., {Gautier}, T.~N., III, {Howell}, S., {Koch}, D.~G.,
  {Latham}, D.~W., {Marcy}, G.~W., {Morehead}, R.~C., and {Sasselov}, D.
  (2011{\natexlab{b}}).
\newblock \enquote{{Architecture and Dynamics of Kepler's Candidate Multiple
  Transiting Planet Systems}.}
\newblock {\em ArXiv e-prints\/}.

\bibitem[{{Maciejewski} et~al.(2010){Maciejewski}, {Dimitrov}, {Neuhaeuser},
  {Niedzielski}, {Raetz}, {Ginski}, {Adam}, {Marka}, {Moualla}, and
  {Mugrauer}}]{Maciejewski10}
{Maciejewski}, G., {Dimitrov}, D., {Neuhaeuser}, R., {Niedzielski}, A.,
  {Raetz}, S., {Ginski}, C., {Adam}, C., {Marka}, C., {Moualla}, M., and
  {Mugrauer}, M. (2010).
\newblock \enquote{{Transit timing variation in exoplanet WASP-3b}.}
\newblock {\em \mnras\/}, 407: 2625--2631.

\bibitem[{{Miller-Ricci} et~al.(2008){Miller-Ricci}, {Rowe}, {Sasselov},
  {Matthews}, {Guenther}, {Kuschnig}, {Moffat}, {Rucinski}, {Walker}, and
  {Weiss}}]{MillerRicci08}
{Miller-Ricci}, E., {Rowe}, J.~F., {Sasselov}, D., {Matthews}, J.~M.,
  {Guenther}, D.~B., {Kuschnig}, R., {Moffat}, A.~F.~J., {Rucinski}, S.~M.,
  {Walker}, G.~A.~H., and {Weiss}, W.~W. (2008).
\newblock \enquote{{MOST Space-based Photometry of the Transiting Exoplanet
  System HD 209458: Transit Timing to Search for Additional Planets}.}
\newblock {\em \apj\/}, 682: 586--592.

\bibitem[{{Moorhead} and {Ford}(2010)}]{Moorhead10}
{Moorhead}, A. and {Ford}, E. (2010).
\newblock \enquote{{Resolving the degeneracy between eccentric planets and 2:1
  mean motion resonances}.}
\newblock In {K.~Go{\.z}dziewski, A.~Niedzielski, \& J.~Schneider} (ed.), {\em
  EAS Publications Series\/}, volume~42 of {\em EAS Publications Series\/},
  161--164.

\bibitem[{{Moorhead} et~al.(2011){Moorhead}, {Ford}, {Morehead}, {Rowe},
  {Borucki}, {Batalha}, {Bryson}, {Caldwell}, {Fabrycky}, {Gautier}, {Koch},
  {Holman}, {Jenkins}, {Li}, {Lissauer}, {Lucas}, {Marcy}, {Quinn}, {Quintana},
  {Ragozzine}, {Shporer}, {Still}, and {Torres}}]{Moorhead11}
{Moorhead}, A.~V., {Ford}, E.~B., {Morehead}, R.~C., {Rowe}, J., {Borucki},
  W.~J., {Batalha}, N.~M., {Bryson}, S.~T., {Caldwell}, D.~A., {Fabrycky},
  D.~C., {Gautier}, T.~N., III, {Koch}, D.~G., {Holman}, M.~J., {Jenkins},
  J.~M., {Li}, J., {Lissauer}, J.~J., {Lucas}, P., {Marcy}, G.~W., {Quinn},
  S.~N., {Quintana}, E., {Ragozzine}, D., {Shporer}, A., {Still}, M., and
  {Torres}, G. (2011).
\newblock \enquote{{The Distribution of Transit Durations for Kepler Planet
  Candidates and Implications for their Orbital Eccentricities}.}
\newblock {\em \apjs\/}.

\bibitem[{{Nesvorn{\'y}}(2009)}]{Nesvorny09}
{Nesvorn{\'y}}, D. (2009).
\newblock \enquote{{Transit Timing Variations for Eccentric and Inclined
  Exoplanets}.}
\newblock {\em \apj\/}, 701: 1116--1122.

\bibitem[{{Nesvorn{\'y}} and {Beaug{\'e}}(2010)}]{Nesvorny10}
{Nesvorn{\'y}}, D. and {Beaug{\'e}}, C. (2010).
\newblock \enquote{{Fast Inversion Method for Determination of Planetary
  Parameters from Transit Timing Variations}.}
\newblock {\em \apjl\/}, 709: L44--L48.

\bibitem[{{Nesvorn{\'y}} and {Morbidelli}(2008)}]{Nesvorny08}
{Nesvorn{\'y}}, D. and {Morbidelli}, A. (2008).
\newblock \enquote{{Mass and Orbit Determination from Transit Timing Variations
  of Exoplanets}.}
\newblock {\em \apj\/}, 688: 636--646.

\bibitem[{{Payne} and {Ford}(2011)}]{Payne11}
{Payne}, M.~J. and {Ford}, E.~B. (2011).
\newblock \enquote{{An Analysis of Jitter and Transit Timing Variations in the
  HAT-P-13 System}.}
\newblock {\em \apj\/}, 729: 98--+.

\bibitem[{{Payne} et~al.(2010){Payne}, {Ford}, and {Veras}}]{Payne10}
{Payne}, M.~J., {Ford}, E.~B., and {Veras}, D. (2010).
\newblock \enquote{{Transit Timing Variations for Inclined and Retrograde
  Exoplanetary Systems}.}
\newblock {\em \apjl\/}, 712: L86--L92.

\bibitem[{{Ragozzine} and {Holman}(2010)}]{Ragozzine10}
{Ragozzine}, D. and {Holman}, M.~J. (2010).
\newblock \enquote{{The Value of Systems with Multiple Transiting Planets}.}
\newblock {\em ArXiv e-prints\/}.

\bibitem[{{Slawson} et~al.(2011){Slawson}, {Prsa}, {Welsh}, {Orosz}, {Rucker},
  {Batalha}, {Doyle}, {Engle}, {Conroy}, {Coughlin}, {Ames Gregg}, {Fetherolf},
  {Short}, {Windmiller}, {Fabrycky}, {Howell}, {Jenkins}, {Uddin}, {Mullally},
  {Seader}, {Thompson}, {Sanderfer}, {Borucki}, and {Koch}}]{Slawson11}
{Slawson}, R.~W., {Prsa}, A., {Welsh}, W.~F., {Orosz}, J.~A., {Rucker}, M.,
  {Batalha}, N.~M., {Doyle}, L.~R., {Engle}, S.~G., {Conroy}, K., {Coughlin},
  J., {Ames Gregg}, T., {Fetherolf}, T., {Short}, D.~R., {Windmiller}, G.,
  {Fabrycky}, D.~C., {Howell}, S.~B., {Jenkins}, J.~M., {Uddin}, K.,
  {Mullally}, F., {Seader}, S.~E., {Thompson}, S.~E., {Sanderfer}, D.~T.,
  {Borucki}, W., and {Koch}, D. (2011).
\newblock \enquote{{Kepler Eclipsing Binary Stars. II. 2165 Eclipsing Binaries
  in the Second Data Release}.}
\newblock {\em ArXiv e-prints\/}.

\bibitem[{{Steffen} and {Agol}(2005)}]{Steffen05}
{Steffen}, J.~H. and {Agol}, E. (2005).
\newblock \enquote{{An analysis of the transit times of TrES-1b}.}
\newblock {\em \mnras\/}, 364: L96--L100.

\bibitem[{{Steffen} et~al.(2010){Steffen}, {Batalha}, {Borucki}, {Buchhave},
  {Caldwell}, {Cochran}, {Endl}, {Fabrycky}, {Fressin}, {Ford}, {Fortney},
  {Haas}, {Holman}, {Isaacson}, {Jenkins}, {Koch}, {Latham}, {Lissauer},
  {Moorhead}, {Morehead}, {Marcy}, {MacQueen}, {Quinn}, {Ragozzine}, {Rowe},
  {Sasselov}, {Seager}, {Torres}, and {Welsh}}]{Steffen10}
{Steffen}, J.~H., {Batalha}, N.~M., {Borucki}, W.~J., {Buchhave}, L.~A.,
  {Caldwell}, D.~A., {Cochran}, W.~D., {Endl}, M., {Fabrycky}, D.~C.,
  {Fressin}, F., {Ford}, E.~B., {Fortney}, J.~J., {Haas}, M.~J., {Holman},
  M.~J., {Isaacson}, H., {Jenkins}, J.~M., {Koch}, D., {Latham}, D.~W.,
  {Lissauer}, J.~J., {Moorhead}, A.~V., {Morehead}, R.~C., {Marcy}, G.,
  {MacQueen}, P.~J., {Quinn}, S.~N., {Ragozzine}, D., {Rowe}, J.~F.,
  {Sasselov}, D.~D., {Seager}, S., {Torres}, G., and {Welsh}, W.~F. (2010).
\newblock \enquote{{Five Kepler target stars that show multiple transiting
  exoplanet candidates}.}
\newblock {\em \apj\/}, 725: 1226--1241.

\bibitem[{{Steffen} et~al.(2011){Steffen}, {Quinn}, {Borucki}, {Brugamyer},
  {Bryson}, {Buchhave}, {Cochran}, {Endl}, {Fabrycky}, {Ford}, {Holman},
  {Jenkins}, {Koch}, {Latham}, {MacQueen}, {Mullally}, {Prsa}, {Ragozzine},
  {Rowe}, {Sanderfer}, {Seader}, {Short}, {Shporer}, {Thompson}, {Torres},
  {Twicken}, {Welsh}, and {Windmiller}}]{Steffen11b}
{Steffen}, J.~H., {Quinn}, S.~N., {Borucki}, W.~J., {Brugamyer}, E., {Bryson},
  S.~T., {Buchhave}, L.~A., {Cochran}, W.~D., {Endl}, M., {Fabrycky}, D.~C.,
  {Ford}, E.~B., {Holman}, M.~J., {Jenkins}, J., {Koch}, D., {Latham}, D.~W.,
  {MacQueen}, P., {Mullally}, F., {Prsa}, A., {Ragozzine}, D., {Rowe}, J.~F.,
  {Sanderfer}, D.~T., {Seader}, S.~E., {Short}, D., {Shporer}, A., {Thompson},
  S.~E., {Torres}, G., {Twicken}, J.~D., {Welsh}, W.~F., and {Windmiller}, G.
  (2011).
\newblock \enquote{{The architecture of the hierarchical triple star KOI 928
  from eclipse timing variations seen in Kepler photometry}.}
\newblock {\em ArXiv e-prints\/}.

\bibitem[{{Veras} et~al.(2011){Veras}, {Ford}, and {Payne}}]{Veras11}
{Veras}, D., {Ford}, E.~B., and {Payne}, M.~J. (2011).
\newblock \enquote{{Quantifying the Challenges of Detecting Unseen Planetary
  Companions with Transit Timing Variations}.}
\newblock {\em \apj\/}, 727: 74--+.

\bibitem[{{Wright}(2005)}]{Wright05}
{Wright}, J.~T. (2005).
\newblock \enquote{{Radial Velocity Jitter in Stars from the California and
  Carnegie Planet Search at Keck Observatory}.}
\newblock {\em \pasp\/}, 117: 657--664.

\bibitem[{{Wright} et~al.(2011){Wright}, {Veras}, {Ford}, {Johnson}, {Marcy},
  {Howard}, {Isaacson}, {Fischer}, {Spronck}, {Anderson}, and
  {Valenti}}]{Wright11}
{Wright}, J.~T., {Veras}, D., {Ford}, E.~B., {Johnson}, J.~A., {Marcy}, G.~W.,
  {Howard}, A.~W., {Isaacson}, H., {Fischer}, D.~A., {Spronck}, J., {Anderson},
  J., and {Valenti}, J. (2011).
\newblock \enquote{{The California Planet Survey. III. A Possible 2:1 Resonance
  in the Exoplanetary Triple System HD 37124}.}
\newblock {\em \apj\/}, 730: 93--+.

\bibitem[{{Youdin}(2011)}]{Youdin11}
{Youdin}, A.~N. (2011).
\newblock \enquote{{The Exoplanet Census: A General Method, Applied to
  Kepler}.}
\newblock {\em ArXiv e-prints\/}.

\end{thebibliography}

\vskip 3ex

\begin{aboutauthors}
Eric B.~ Ford is an associate professor of astronomy at the University of Florida.  His research focuses on extrasolar planets, including both theory and the statistical interpretation of exoplanet observations.
Althea V. Moorhead is a postdoctoral associate studying planet formation and contributing to NASA's {\em Kepler} mission.  Dr. Moorhead applied the surrogate model to real and simulated radial velocity observations.  
Dimitri Veras was a postdoctoral associate studying planetary orbital dynamics and the transit timing method.  Dr. Veras applied the surrogate model to simulated transit timing variations data sets.  Dr. Veras is now a postdoctoral associate at the Institute of Astronomy in Cambridge, UK.  
\end{aboutauthors}

\begin{acknowledgement}
The authors wish to thank Malay Ghosh, Phil Gregory, Tom Loredo and Matthew Payne for valuable discussions and feedback.  
This material is based upon work supported by the National Science Foundation under Grant No. 0707203.  
Additional support for this work was provided by the National Aeronautics and Space Administration under grant NNX09AB35Gb issued through the Origins of Solar Systems program and grant NNX08AR04G issued through the Kepler Participating Scientist Program.  
\end{acknowledgement}

\

\end{document}